\documentclass[preprint,review,12pt]{elsarticle}

\usepackage{amssymb}
\usepackage{graphicx}

\journal{J. Alloys Comp.}

\begin{document}

\begin{frontmatter}

\title{Synthesis and physical properties of a new caged compound Ce$_{3}$Pd$_{20}$As$_{6}$ of the C$_{6}$Cr$_{23}$-type structure}

\author{K. Orita$^1$}

\address{$^1$ Department of Electrical Engineering, Faculty of Engineering, Fukuoka Institute of Technology, 3-30-1 Wajiro-higashi, Higashi-ku, Fukuoka 811-0295, Japan}

\author{K. Uenishi$^2$}

\address{$^2$ Department of Quantum Matter, ADSM, Hiroshima University, 1-3-1 Kagamiyama, Higashi-Hiroshima 739-8530, Japan}

\author{M. Tsubota$^3$}

\address{$^3$ Physonit Inc., 6-10 Minami-Horikawa, Kaita Aki, Hiroshima 736-0044, Japan}

\author{Y. Shimada$^2$}

\address{$^2$ Department of Quantum Matter, ADSM, Hiroshima University, 1-3-1 Kagamiyama, Higashi-Hiroshima 739-8530, Japan}

\author{T. Onimaru$^2$}

\address{$^2$ Department of Quantum Matter, ADSM, Hiroshima University, 1-3-1 Kagamiyama, Higashi-Hiroshima 739-8530, Japan}

\author{T. Takabatake$^2$}

\address{$^2$ Department of Quantum Matter, ADSM, Hiroshima University, 1-3-1 Kagamiyama, Higashi-Hiroshima 739-8530, Japan}

\author{J. Kitagawa$^1$}

\address{$^1$ Department of Electrical Engineering, Faculty of Engineering, Fukuoka Institute of Technology, 3-30-1 Wajiro-higashi, Higashi-ku, Fukuoka 811-0295, Japan}
\ead{j-kitagawa@fit.ac.jp}

\author{}

\address{}

\author{}

\address{}

\begin{abstract}
We have found that Ce$_{3}$Pd$_{20}$As$_{6}$ crystallizes into a cubic C$_{6}$Cr$_{23}$-type structure. Combination of electron probe microanalysis  of the chemical composition and Rietveld analysis of the powder X-ray diffraction pattern has revealed an inhomogeneous atomic composition of variable stoichiometry. The physical properties of Ce$_{3}$Pd$_{20}$As$_{6}$ were investigated by measuring the magnetization, electrical resistivity and specific heat. The 4$f$ electrons of Ce$^{3+}$ ions are well localized but do not show phase transition down to 0.5 K. The metallic electrical resistivity shows a weak Kondo screening. The residual resistivity ratio is rather low probably due to the variable stoichiometry. The magnetization curve and magnetic entropy suggest the $\Gamma_{8}$ quartet crystalline-electric-field ground state at least one of two Ce sites.
\end{abstract}


\end{frontmatter}

\clearpage

\section{Introduction}
The intermetallic compounds with the cubic C$_{6}$Cr$_{23}$-type structure (space group: Fm$\bar{3}$m) are attracting considerable attention as ferromagnetic materials\cite{Buschow:JMMM1983,Eriksson:JMMM2007}.
Several R$_{3}$Pd$_{20}$(Si or Ge)$_{6}$ (R:rare earth) compounds with the C$_{6}$Cr$_{23}$-type structure show successive antiferromagnetic orderings at two different R sites\cite{Hermann:JPCM1999,Donni:JALCOM2000}.
Another interesting research area of the C$_{6}$Cr$_{23}$-type structure is the multipolar effect associated with the orbital degeneracy of 4$f$ (5$f$) electrons in lanthanide (actinide)-based systems.
The degeneracy of $J$ multiplet of the lanthanide (actinide) ion is lifted under a crystalline-electric-field (CEF) produced by ligand atoms.
The CEF ground state often possesses the orbital degeneracy, which can be lifted by multipolar orderings, such as quadrupolar and octupolar orderings\cite{Santini:RMP2009}.

We have focused on a series of ternary compounds of Ce-Pd-X (X=Si, Ge and P) with the C$_{6}$Cr$_{23}$-type structure\cite{Takeda:JPSP1995,Kitagawa:PRB1996,Kitagawa:JAC1997,Kitagawa:PRB1998,Kitagawa:JPSJ2000,Nemoto:PRB2003,Goto:JPSJ2009,Deen:PRB2010,Custers:NM2012,Ono:JPCM2013,Abe:RP2014}, because this system provides a unique opportunity of systematic investigation of multipolar effects.
These compounds have two crystallographically-inequivalent Ce-sites of the 4{\it a} ($O_{h}$ symmetry) and 8{\it c} ($T_{d}$ symmetry) sites.
The six-fold degeneracy of the $J$=5/2 multiplet of the Ce$^{3+}$ ion is lifted into the $\Gamma_{7}$ doublet and $\Gamma_{8}$ quartet under a cubic CEF.
The CEF ground states of the compounds of X=Si and Ge are the $\Gamma_{8}$ states at both Ce sites\cite{Kitagawa:PRB1996,Kitagawa:JPSJ2000,Kitazawa:PC}, which contribute to the quadrupolar and magnetic orderings.
Each compound (X=Si or Ge) shows the coexistence of the Kondo-lattice behavior and ordered states\cite{Takeda:JPSP1995,Kitagawa:PRB1996,Kitagawa:JPSJ2000,Nemoto:PRB2003}.
On the other hand, the different CEF ground state at either 4{\it a} or 8{\it c} site is possibly realized in the phosphide with atomic disorder at the 4{\it a} site, where the Ce atoms are partially replaced with Pd atoms\cite{Abe:RP2014}.
In the phosphide, the localized 4$f$ electrons show neither pronounced Kondo effect nor long-range ordered states down to 0.5 K\cite{Abe:RP2014}.

We tried replacing the P atom with the As atom of the isoelectronic series of element, expecting a new compound showing multipolar effects.
We have found that Ce$_{3}$Pd$_{20}$As$_{6}$ is a new member of the C$_{6}$Cr$_{23}$-type structure.
In this paper, we report the synthesis and characterization of polycrystalline samples.
The magnetic and transport properties of Ce$_{3}$Pd$_{20}$As$_{6}$ were investigated by measuring the magnetization, electrical resistivity and specific heat.

\section{Experimental}
Polycrystalline samples were prepared using Ce pieces (99.9\%), Pd shot, powder, or sponge (99.9\%) and As powder or grains (99.9\%).
The starting materials of all samples are summarized in Table 1.
Two preparation techniques were employed.
One was the solid-state reaction of CePd$_{3}$, Pd and As.
CePd$_{3}$ was prepared by arc melting the constituent elements (Ce pieces and Pd shot) in an argon atmosphere.
Finely ground CePd$_{3}$, Pd and As with the molar ratio of Ce:Pd:As=3:20:6 were homogeneously mixed together.
The pelletized sample was reacted in an evacuated quartz tube at 900 $^{\circ}$C for 3 days.
The other method was a direct reaction of a pelletized mixture of Ce pieces, Pd sponge and As grains.
The pellet in a carbon crucible was placed in an evacuated quartz tube, heated to 400$^{\circ}$C, held at that temperature for 16 h and further heated to 900 $^{\circ}$C (or 1000$^{\circ}$C), held at that temperature for 10 h.
The reference compound La$_{3}$Pd$_{20}$As$_{6}$ was also synthesized by the direct reaction of La pieces, Pd sponge and As grains at 900 $^{\circ}$C, followed by sintering at 900 $^{\circ}$C for 3 days after regrinding.
All products were characterized using the powder X-ray diffraction (XRD) analysis.
Wave-length dispersive electron probe microanalysis (EPMA) was performed to identify the composition of sample \#4.

\begin{table}
\caption{Starting materials and lattice parameters of prepared samples.}
\label{t1}
\begin{tabular}{ccccc}
\hline
sample No. & CePd$_{3}$ or Ce & Pd & As & {\it a} (\AA) \\
\hline
\#1 & CePd$_{3}$ & powder & powder & 12.361(2) \\
\#2 & CePd$_{3}$ & powder & grain  & 12.435(1) \\
\#3 & CePd$_{3}$ & sponge & grain  & 12.449(4) \\
\#4 & Ce & sponge & grain  & 12.440(2) \\
\#5 & Ce & sponge & grain  & 12.448(3) \\
\hline
\end{tabular}
\end{table}

The temperature dependence of the DC magnetization $\chi$(T) from 1.8 K to 350 K under a magnetic field of 0.1 T was measured using a Quantum Design MPMS.
The magnetization curve was also measured up to 5 T at several temperatures.
The temperature dependence of electrical resistivity $\rho$(T) from 2.5 K to 300 K was measured by an AC four-probe method using a GM refrigerator.
The specific heat $C_{p}$ data from 0.5 K to 300 K was obtained using a Quantum Design PPMS.

\begin{figure}
\begin{center}
\includegraphics[width=8cm]{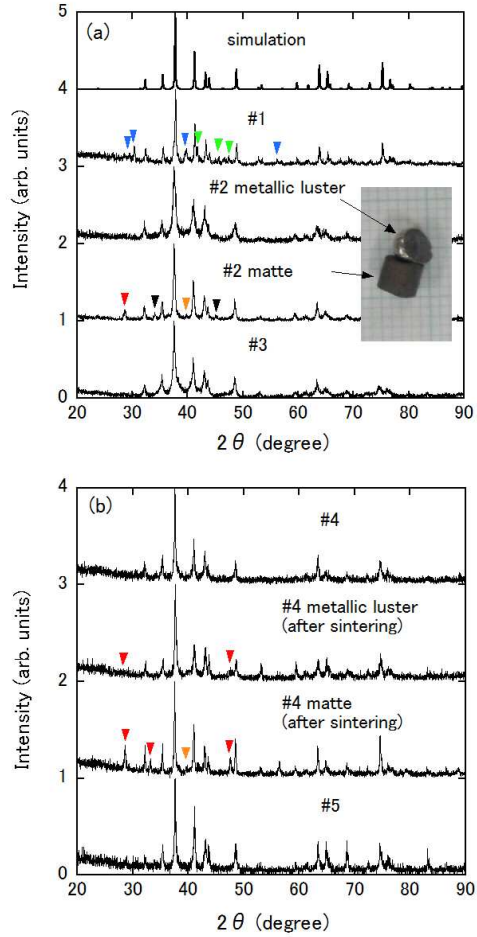}
\end{center}
\caption{(a) XRD patterns of Ce$_{3}$Pd$_{20}$As$_{6}$ \#1, \#2 and \#3 synthesized by solid-state reaction technique. The simulated pattern of C$_{6}$Cr$_{23}$-type structure is also shown. The origin of each pattern is shifted by an integer value for clarity. The inset is the photograph of sample \#2 showing the segregation. (b) XRD patterns of Ce$_{3}$Pd$_{20}$As$_{6}$ \#4 and \#5 synthesized by the direct reaction technique. For the sample \#4, the XRD patterns are presented for metallic luster and matte parts obtained after the sintering. The origin of each pattern is shifted by an integer value for clarity. The blue, green, red, orange and black triangles show Ce$_{2}$O$_{3}$, Pd$_{3}$As, CeO$_{2}$, CePd$_{5}$ and unknown binary (ternary) phases, respectively.}
\label{f1}
\end{figure}

\section{Results and Discussion}
The XRD patterns of samples \#1, \#2 and \#3 prepared by the solid-state reaction technique are shown in Fig.\ 1(a).
All patterns, except for the peaks denoted by triangles, can be indexed by the cubic space group Fm$\bar{3}$m.
The simulated pattern in Fig.\ 1(a) is calculated by using the atomic position of As identical to Si in Ce$_{3}$Pd$_{20}$Si$_{6}$ and closely matches the experimental patterns\cite{Grivanov:JAC1994}.
The sample \#1 using As powder as the starting material shows noticeable impurity peaks (see triangles).
It is unique for sample \#2 that the segregation occurs during the reaction as shown in the inset of Fig.\ 1(a).
The both metallic luster and matte parts form the C$_{6}$Cr$_{23}$-type structure.
The former is almost single phase, but the diffraction peaks are rather broad.
On the contrary, the latter shows the sharp XRD peaks, and impurity peaks denoted by triangles.
The segregation does not occur in sample \#3 synthesized by Pd sponge and As grains.
Although the sample \#3 has little amounts of impurity phases, the XRD peaks are highly broadened.
Consequently, we gave up making good samples based on the solid-state reaction technique.

Figure 1(b) shows the XRD patterns of samples \#4 and \#5 synthesized by the direct-reaction technique.
The direct reaction of constituent elements at 900 $^{\circ}C$ (\#4) or 1000 $^{\circ}C$ (\#5) results in the single-phased C$_{6}$Cr$_{23}$-type structure (see the top and bottom patterns).
The XRD peaks are sharper than those of samples \#2 (metallic luster) and \#3.
However, for sample \#4, after crushing and homogeneous mixing followed by the sintering at 900 $^{\circ}C$ for 3 days, the segregation is observed, which is similar to the results of sample \#2 synthesis.
The XRD pattern of matte part \#4 shows sharper peaks, but includes impurity peaks (see triangles).
The metallic luster part \#4 is almost single phase with small amount of impurity phase CeO$_{2}$.
Because the metallic luster part \#4 is nearly single phase and has the rather sharp XRD peaks, we have carried out the EPMA examination and investigation of physical properties of this part.
The lattice parameters of all prepared samples were calculated by the least mean square method and given in Table 1.

\begin{figure}
\begin{center}
\includegraphics[width=9cm]{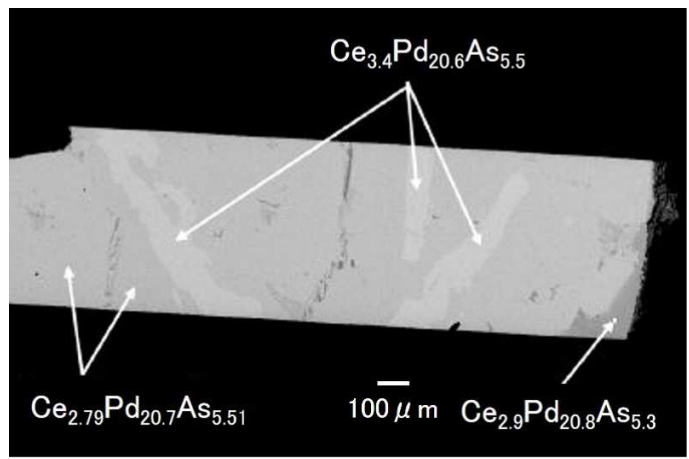}
\end{center}
\caption{Back-scattered electron (15 keV) image of Ce$_{3}$Pd$_{20}$As$_{6}$ \#4 (metallic luster).}
\label{f2}
\end{figure}

The back-scattered electron image of metallic luster part \#4 obtained by EPMA with electron beams of 15 keV is shown in Fig.\ 2, where the three different-contrasted-images are observed.
The atomic composition of each phase was found to be different, maintaining 116 atoms in the unit cell (defect free C$_{6}$Cr$_{23}$-type structure) as denoted in Fig.\ 2.
This hypothesis was verified by checking that the three phases could reproduce the XRD pattern, with the help of the Rietveld analysis as mentioned below.

\begin{figure}
\begin{center}
\includegraphics[width=13cm]{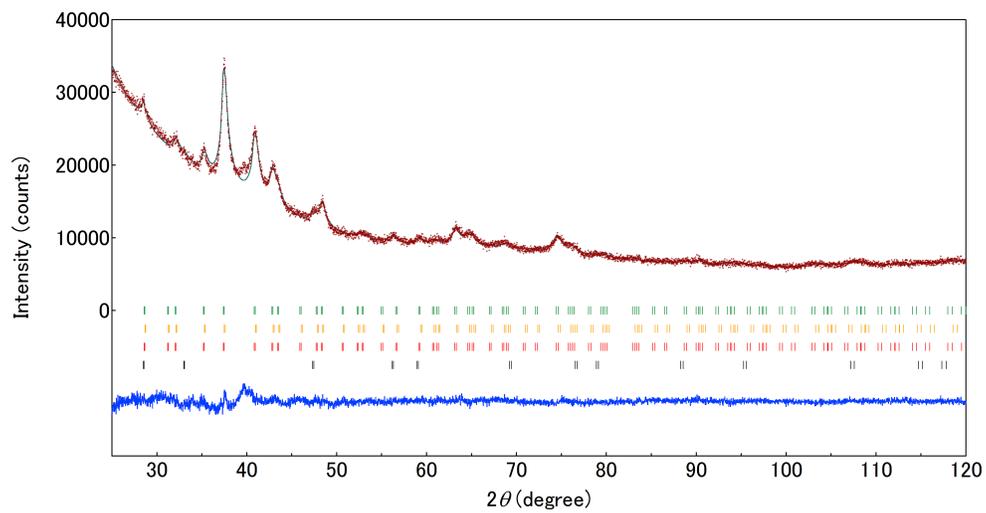}
\end{center}
\caption{Rietveld refinement of Ce$_{3}$Pd$_{20}$As$_{6}$ \#4 (metallic luster). The observed (+) and calculated (solid line) XRD patterns are shown at the top. The difference between the observed and calculated XRD patterns is shown at the bottom. The four sets of tick marks indicate the positions of Bragg reflections for Ce$_{2.79}$Pd$_{20.7}$As$_{5.51}$, Ce$_{3.4}$Pd$_{20.6}$As$_{5}$, Ce$_{2.9}$Pd$_{20.8}$As$_{5.3}$ and CeO$_{2}$ from top to bottom.}
\label{f3}
\end{figure}

The XRD pattern of sample \#4 (metallic luster part) was fitted by the Rietveld refinement program RIETAN-FP\cite{Izumi:SSP2007}.
Figure 3 shows the refinement result ($R_{wp}$$=$3.18\% and $R_{p}$$=$2.47\%), which gives 93 wt\% of the main phase and 7 wt\% CeO$_{2}$ as an additional phase.
The main phase is assumed to be composed of 75.4 wt\% Ce$_{2.79}$Pd$_{20.7}$As$_{5.51}$ (A phase), 16.7 wt\% Ce$_{3.4}$Pd$_{20.6}$As$_{5}$ (B phase) and 0.83 wt\% Ce$_{2.9}$Pd$_{20.8}$As$_{5.3}$ (C phase), deduced from EPMA.
The lattice parameters, atomic coordinates, displacement parameters and site occupancies for the A, B and C phases are listed in Table 2.
In the A and C phases, the Ce atoms at the 4{\it a} site and the As atoms at the 24{\it e} site are partially replaced with Pd atoms.
The B phase possesses the atomic disorder of As, Ce and Pd atoms at only the 24{\it e} site.
Recent our research on Ce$_{3}$Pd$_{20}$P$_{6}$ has revealed the atomic disorder at the 4a site of the C$_{6}$Cr$_{23}$-type structure\cite{Abe:RP2014}.
Therefore it is characteristic of the C$_{6}$Cr$_{23}$-type Ce-Pd-X (X=P and As) compounds that the phases are stabilized by the inevitable disorder, which is contrasted with negligible atomic disorder\cite{Grivanov:JAC1994} in Ce$_{3}$Pd$_{20}$Si$_{6}$ and Ce$_{3}$Pd$_{20}$Ge$_{6}$.
The atomic composition averaged over the A, B and C phases is Ce$_{2.9}$Pd$_{20.7}$As$_{5.4}$, which is close to Ce$_{3}$Pd$_{20}$As$_{6}$.
Therefore, we hereafter keep on denoting Ce$_{3}$Pd$_{20}$As$_{6}$ as the composition of sample \#4 (metallic luster).

\begin{table}
\caption{Atomic coordinates, equivalent isotropic displacement parameters and site occupancies for the three distinguished parts of the sample Ce$_{3}$Pd$_{20}$As$_{6}$ \#4 (metallic luster). The estimated volume fraction of each phase is also shown.}
\label{t2}
\footnotesize
\begin{tabular}{ccccccc}
\hline
atom & site & $x$ & $y$ & $z$ & $U_{eq}$ (\AA$^{2}$$\times$10$^{2}$) & occupancy \\
\hline
\multicolumn{7}{c}{A: Ce$_{2.79}$Pd$_{20.7}$As$_{5.51}$ ($a$=12.440(2)\AA), 81 vol\%} \\
Ce(1) & 4{\it a} & 0 & 0 & 0 & 1.84(1) & 0.79  \\
Pd(3) & 4{\it a} & 0 & 0 & 0 & 1.84(1) & 0.21  \\
Ce(2) & 8{\it c} & 0.25 & 0.25 & 0.25 & 1.84(1) & 1  \\
Pd(2) & 32{\it f} & 0.3821(1) & 0.3821(1) & 0.3821(1) & 1.44(1) & 1  \\
Pd(1) & 48{\it h} & 0 & 0.1819(1) & 0.1819(1) & 1.44(1) & 1  \\
As & 24{\it e} & 0.2590(1) & 0 & 0 & 0.16(1) & 0.92  \\
Pd(4) & 24{\it e} & 0.2590(1) & 0 & 0 & 0.16(1) & 0.08  \\
&&&&&& \\
\multicolumn{7}{c}{B: Ce$_{3.4}$Pd$_{20.6}$As$_{5}$ ($a$=12.404(8)\AA), 18 vol\%} \\
Ce(1) & 4{\it a} & 0 & 0 & 0 & 1.84(1) & 0.79  \\
Ce(2) & 8{\it c} & 0.25 & 0.25 & 0.25 & 1.84(1) & 1  \\
Pd(2) & 32{\it f} & 0.3821(1) & 0.3821(1) & 0.3821(1) & 1.44(1) & 1  \\
Pd(1) & 48{\it h} & 0 & 0.1819(1) & 0.1819(1) & 1.44(1) & 1  \\
As & 24{\it e} & 0.2590(1) & 0 & 0 & 0.16(1) & 0.83  \\
Ce(3) & 24{\it e} & 0.2590(1) & 0 & 0 & 0.16(1) & 0.07  \\
Pd(3) & 24{\it e} & 0.2590(1) & 0 & 0 & 0.16(1) & 0.1  \\
&&&&&& \\
\multicolumn{7}{c}{C: Ce$_{2.9}$Pd$_{20.8}$As$_{5.3}$ ($a$=12.440(2)\AA), 1 vol\%} \\
Ce(1) & 4{\it a} & 0 & 0 & 0 & 1.84(1) & 0.9  \\
Pd(3) & 4{\it a} & 0 & 0 & 0 & 1.84(1) & 0.1  \\
Ce(2) & 8{\it c} & 0.25 & 0.25 & 0.25 & 1.84(1) & 1  \\
Pd(2) & 32{\it f} & 0.3821(1) & 0.3821(1) & 0.3821(1) & 1.44(1) & 1  \\
Pd(1) & 48{\it h} & 0 & 0.1819(1) & 0.1819(1) & 1.44(1) & 1  \\
As & 24{\it e} & 0.2590(1) & 0 & 0 & 0.16(1) & 0.883  \\
Pd(4) & 24{\it e} & 0.2590(1) & 0 & 0 & 0.16(1) & 0.117  \\
\hline
\end{tabular}
\end{table}

\begin{figure}
\begin{center}
\includegraphics[width=10cm]{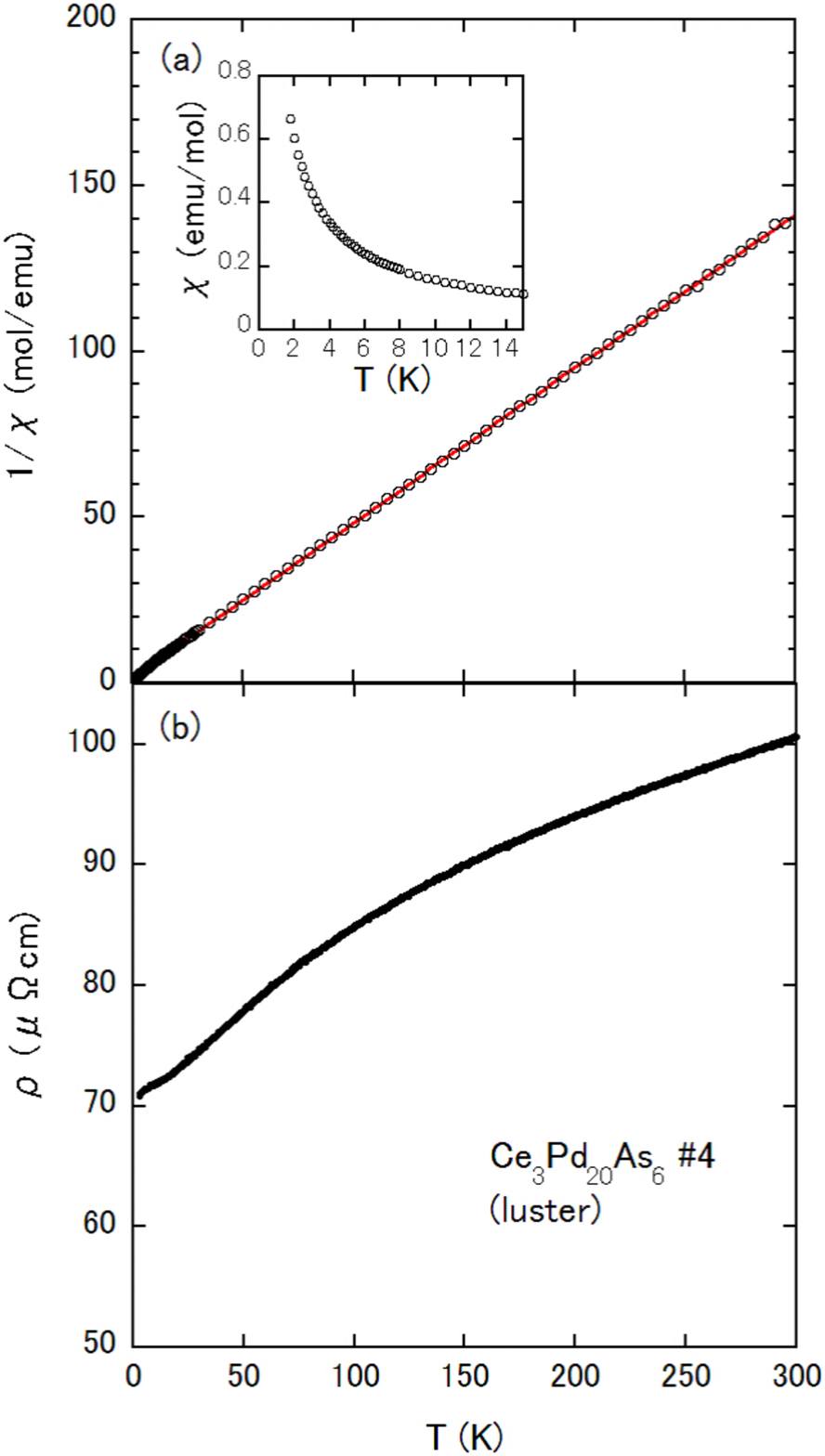}
\end{center}
\caption{Temperature dependences of (a) inverse magnetic susceptibility and (b) electrical resistivity of Ce$_{3}$Pd$_{20}$As$_{6}$ \#4 (metallic luster). The inset shows the low-temperature $\chi$ of the sample.}
\label{f4}
\end{figure}

Figure 4(a) shows the temperature dependence of the reciprocal $\chi$ of Ce$_{3}$Pd$_{20}$As$_{6}$ \#4 (metallic luster).
It follows the Curie-Weiss law above 100 K (see the red line in Fig.\ 4(a)).
The effective moment $\mu_{eff}$ and Weiss temperature $\Theta$ are 2.43 $\mu_{B}$/Ce and -3.3 K, respectively.
The value of $\mu_{eff}$ is slightly smaller than 2.54 $\mu_{B}$/Ce, expected for a free trivalent Ce ion, presumably due to the existence of impurity CeO$_{2}$ phase with tetravalent Ce ions.
The inset of Fig.\ 4(a) is $\chi$(T) below 15 K, which exhibits no magnetic ordering down to 1.8 K.

The data of $\rho$(T) of Ce$_{3}$Pd$_{20}$As$_{6}$ \#4 (metallic luster) indicates a metallic behavior (see Fig.\ 4(b)).
Contrary to the Kondo-lattice behavior in the compounds\cite{Takeda:JPSP1995,Kitagawa:PRB1996} of X=Si and Ge, Ce$_{3}$Pd$_{20}$As$_{6}$ shows no Kondo effect, suggesting a relatively weak Kondo screening.
The residual resistivity ratio is rather low, which is attributable to the inhomogeneity of atomic composition with variable stoichiometry.

\begin{figure}
\begin{center}
\includegraphics[width=9cm]{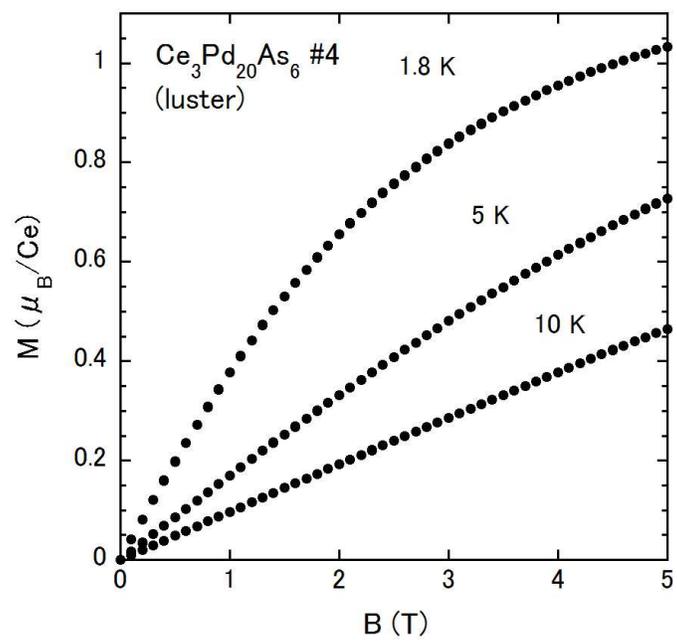}
\end{center}
\caption{Magnetization curves of Ce$_{3}$Pd$_{20}$As$_{6}$ \#4 (metallic luster) measured at 1.8, 5 and 10 K.}
\label{f5}
\end{figure}

As shown in Fig.\ 5, $M$(B $=$ 5 T) at 1.8 K of Ce$_{3}$Pd$_{20}$As$_{6}$ \#4 (metallic luster) exceeds the value of 0.714 $\mu_{B}$/Ce expected for the saturation moment of the isolated $\Gamma_{7}$ state.
Because the two crystallographically inequivalent Ce sites of the 4{\it a} and 8{\it c} sites exist, we have to assign CEF ground states for both Ce sites.
The result of magnetization curve suggests that at least one of the two is $\Gamma_{8}$.

\begin{figure}
\begin{center}
\includegraphics[width=10cm]{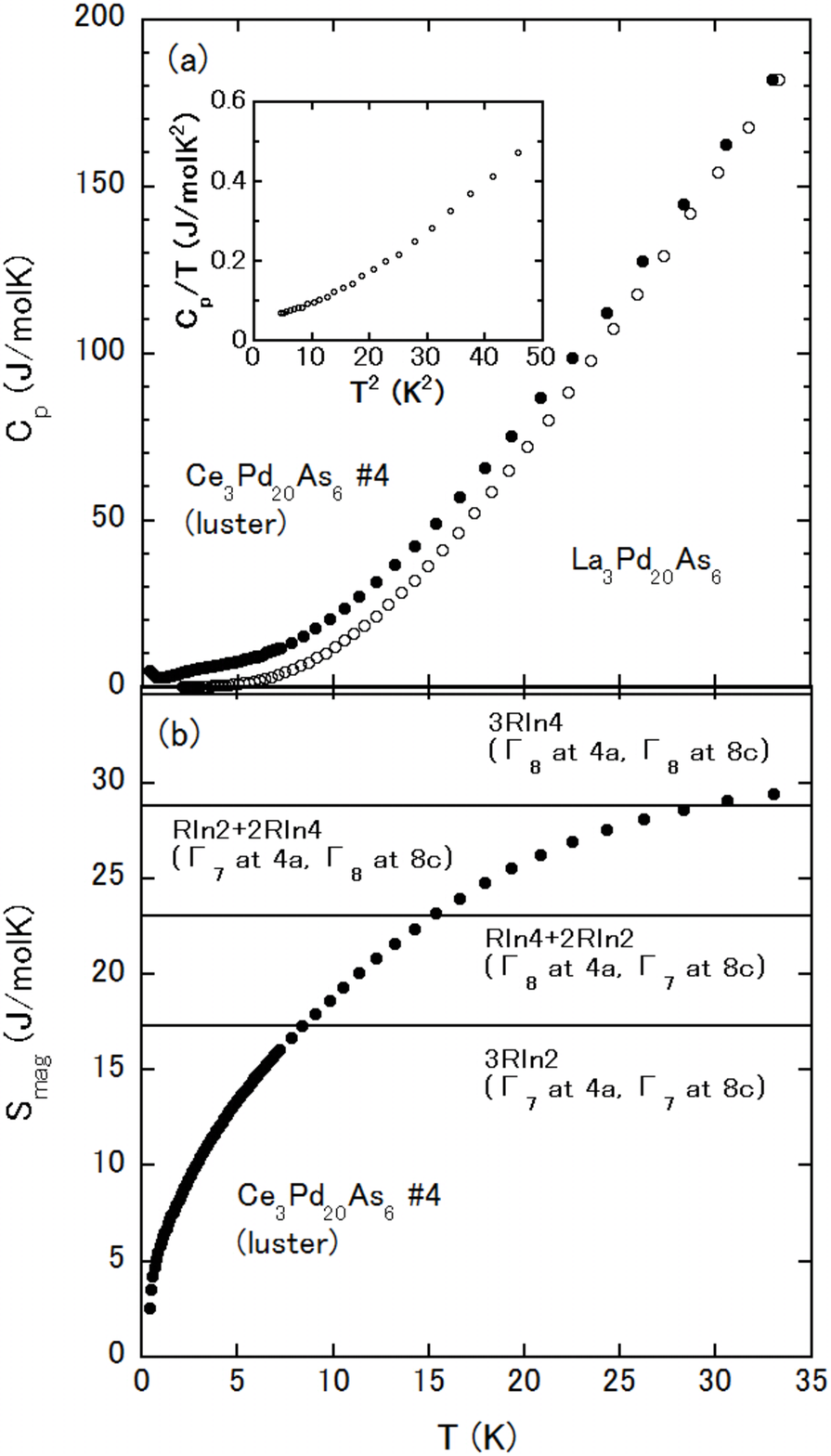}
\end{center}
\caption{Temperature dependences of (a) specific heat and (b) magnetic entropy of Ce$_{3}$Pd$_{20}$As$_{6}$ \#4 (metallic luster). $C_{p}$(T) of La$_{3}$Pd$_{20}$As$_{6}$ is also shown in (a). The inset of (a) is $C_{p}/T$ vs $T^{2}$ plot of La$_{3}$Pd$_{20}$As$_{6}$.}
\label{f6}
\end{figure}

The temperature dependences of specific heat $C_{p}$(T) of Ce$_{3}$Pd$_{20}$As$_{6}$ \#4 (metallic luster) and the La counterpart are shown in Fig.\ 6(a).
$C_{p}$(T) of Ce$_{3}$Pd$_{20}$As$_{6}$ shows no pronounced phase transition down to 0.5 K, while small upturn below 0.9 K implies the possible occurrence of phase transition below 0.5 K.
The inset of Fig.\ 6(a) is the $C_{p}/T$ vs $T^{2}$ plot for La$_{3}$Pd$_{20}$As$_{6}$.
Below about 3 K, $C_{p}/T$ follows $\gamma + \beta T^{2}$ in which the electronic specific heat coefficient $\gamma=$ 47 mJ/molK$^{2}$ and $\beta=$ 4.86 mJ/molK$^{4}$ (the Debye temperature\cite{remark} $=$ 226 K).

We calculated the temperature dependence of magnetic entropy $S_{mag}$(T) by integrating $C_{mag}/T$ with respect to the temperature (see Fig.\ 6(b)).
The magnetic part $C_{mag}$ was obtained by subtracting $C_{p}$(T) of La$_{3}$Pd$_{20}$As$_{6}$, which was assumed to follow $\gamma T + \beta T^{3}$ below 2 K, from that of the Ce counterpart.
$C_{mag}/T$ below 0.5 K was linearly extrapolated to zero at 0 K.
Since $C_{mag}/T$ at 0 K, corresponding to the electronic specific heat coefficient of 4$f$ electrons, should be nonzero, the actual $S_{mag}$(T) would be larger than that in Fig.\ 6(b).
Therein, we denote the entropy values associated with the combinations of CEF ground states at the 4{\it a} and 8{\it c} sites.
As the temperature is increased, $S_{mag}$ exceeds 3Rln2, at 10 K, and approaches the entropy value being ascribable to the $\Gamma_{7}$ ($\Gamma_{8}$) ground state at the 4{\it a} (8{\it c}) site.
This behavior in $S_{mag}$(T) suggests the $\Gamma_{8}$ ground state at either 4{\it a} or 8{\it c} site, which is consistent with the result of magnetization curve.

The possible CEF ground states of the $\Gamma_{7}$ and $\Gamma_{8}$ states depending on the Ce sites for Ce$_{3}$Pd$_{20}$As$_{6}$ is similar to the case of the phosphide\cite{Abe:RP2014}.
This is in contrast with $\Gamma_{8}$ ground states at both Ce sites in Ce$_{3}$Pd$_{20}$X$_{6}$ (X=Si and Ge). 
Based on the crystallographic consideration, we discuss why the ground states are different between Ce$_{3}$Pd$_{20}$As$_{6}$ (and the phosphide) and Ce$_{3}$Pd$_{20}$X$_{6}$ (X=Si and Ge).
The Ce atom at the 4{\it a} site is caged in a polyhedron composed of 6 X (= Si, Ge, P, or As) atoms and 12 Pd atoms, whereas the one at the 8{\it c} site in a polyhedron composed of 16 Pd atoms.
The CEF originating from P or As atom thus affects the Ce 4$f$ state at the 4{\it a} site, and may allow the $\Gamma_{7}$ ground state.
On the other hand, the Ce atom at the 8{\it c} site is surrounded by only Pd atoms.
So the CEF ground state at the 8{\it c} site would not depend on the X atom, maintaining the $\Gamma_{8}$ state even in Ce$_{3}$Pd$_{20}$As$_{6}$ (and the phosphide).
Therefore the $\Gamma_{7}$ and $\Gamma_{8}$ ground states are expected at the 4{\it a} and 8{\it c} sites, respectively, in Ce$_{3}$Pd$_{20}$As$_{6}$.

\section{Summary}

We have found Ce$_{3}$Pd$_{20}$As$_{6}$ crystallizing into the cubic C$_{6}$Cr$_{23}$-type structure.
During the reaction of constituent elements, we have often observed the segregation, which suggests the unstable crystal structure of Ce$_{3}$Pd$_{20}$As$_{6}$.
In fact, the results of EPMA and Rietveld analysis indicate the inhomogeneity of atomic composition of variable stoichiometry.
The physical properties were investigated by measuring $\chi$, $\rho$ and $C_{p}$.
The residual resistivity ratio of $\rho$(T) showing metallic behavior is rather low, which is ascribed to the inhomogeneous atomic composition with variable stoichiometry. 
The localized Ce moments show neither pronounced Kondo effect nor phase transitions down to 0.5 K.
At least one of two CEF ground states is the $\Gamma_{8}$ quartet, which is deduced from the analysis of magnetization curve and magnetic entropy.

\section*{Acknowledgments}
We thank Y. Shibata for the electron probe microanalysis.
J.K. is grateful for the financial support provided by the Asahi Glass Foundation and Comprehensive Research Organization of Fukuoka Institute of Technology.


\begin{thebibliography}{99}
\bibitem{remark}
If we consider acoustic and optical phonon modes, which can be described by the Debye and Einstein models, respectively, a lower Debye temperature is obtained. However, in this report, for the comparison with reported Debye temperatures of the isostructural La$_{3}$Pd$_{20}$X$_{6}$ (X=Si or Ge) compounds, employing the Debye model taking into account all phonon modes, we give the value obtained by the same model.
\end{thebibliography}
\end{document}